# Hydrated Excess Protons in Acetonitrile/Water Mixtures – Solvation Species and Ultrafast Proton Motions

Achintya Kundu[1], Fabian Dahms[1], Benjamin P. Fingerhut[1], Erik T. J. Nibbering[1], Ehud Pines[2], Thomas Elsaesser[1*]

[1]*Max-Born-Institut für Nichtlineare Optik und Kurzzeitspektroskopie, Berlin, 12489, Germany*

[2]*Department of Chemistry, Ben Gurion University of the Negev, Beer-Sheva 84105, Israel*

*Corresponding author. Email address elsasser@mbi-berlin.de

Abstract

The solvation structure of protons in aqueous media is highly relevant to electric properties and to proton transport in liquids and membranes. At ambient temperature, polar liquids display structural fluctuations on femto- to picosecond time scales with a direct impact on proton solvation. We apply two-dimensional infrared (2D-IR) spectroscopy for following proton dynamics in acetonitrile/water mixtures with the Zundel cation $H_5O_2^+$ prepared in neat acetonitrile as a benchmark. The 2D-IR spectra of the proton transfer mode of $H_5O_2^+$ demonstrate stochastic large-amplitude motions in the double-minimum proton potential, driven by fluctuating electric fields. In all cases the excess proton is embedded in a water dimer, forming an $H_5O_2^+$ complex as major solvation species. This observation is rationalized by quantum mechanics/molecular mechanics molecular dynamics simulations including up to 4 water molecules embedded in acetonitrile. The Zundel motif interacts with its closest water neighbor in an $H_7O_3^+$ unit without persistent proton localization.

**TOC figure**

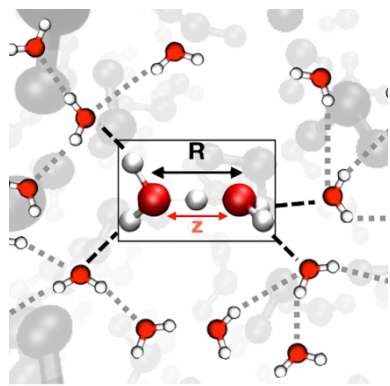

The hydration of ions plays a fundamental role in liquid-phase chemistry, both for molecular structures and chemical processes.[1-3] Properties of solvated ions have been studied with a broad range of experimental methods, addressing both time-averaged equilibrium geometries and structure changes on time scales down to femtoseconds. Such experimental work has been complemented by theoretical calculations of intermolecular interactions and simulations of molecular dynamics. In this context, the solvation of protons in water and other polar liquids represents a prototypical case, playing a fundamental role for proton transport in liquids, acids[4-6], biological and artificial membranes[7,8], and energy storage devices such as batteries and fuel cells.[9-11] The characterization of the main proton solvation structures, their dynamics and associated lifetimes represents a major prerequisite for understanding such phenomena at the molecular level.

There are two limiting solvation structures of hydrated excess protons in water: (i) the Zundel cation[12] $H_5O_2^+$ accommodates a proton $H^+$ between two flanking water molecules with the proton transfer coordinate $z = r_{O1\cdots H+} - r_{O2\cdots H+}$ (cf. Fig. 1a; $r_{O1\cdots H+}$, $r_{O2\cdots H+}$: distances between the proton $H^+$ and the oxygen atoms of the flanking water molecules 1 and 2) and the $O\cdots O$ distance (hydrogen bond coordinate) $R$. In the limiting geometry of a symmetric Zundel cation the proton resides at equal distances to the two oxygen atoms ($z = 0$). The Zundel geometry persists as long as the proton is confined by an $O\cdots O$ double minimum potential with low double-well barrier. (ii) the Eigen complex $H_9O_4^+$ (Fig. 1c) in which a central hydronium $H_3O^+$ forms hydrogen bonds with three neighboring water molecules and the proton is localized at the hydronium core.[13] Stationary Zundel, Eigen, and intermediate proton solvation structures have been demonstrated in ultracold protonated water clusters[14-18] where fluctuations of the hydration geometries are frozen out.

In contrast, protons in a liquid water environment at ambient temperature are subject to fluctuating electric forces with local fluctuation amplitudes of some 20-30 MV/cm, caused by thermally excited structure fluctuations of the water dipoles.[19,20] Such structure fluctuations cover a time range from tens of femtoseconds up to several picoseconds[21] with the fastest motions of water molecules originating from librational excitations.[22] Hydrogen bond breaking and reformation between water molecules are clocked in picoseconds.[23]

Structure and dynamics of protons in this highly complex liquid environment have been the subject of extensive theoretical and simulation work with partly conflicting results.[4-6,24-29] In particular, the relative abundance of Zundel, Eigen, and/or intermediate solvation species in neat water has remained controversial. On the experimental side, femtosecond infrared spectroscopy has probed the fluctuating ensemble by vibrational excitations of water molecules.[30-32] Fluctuating forces manifest in transient vibrational lineshapes, most specifically in two-dimensional infrared (2D-IR) spectra. For protons in water, the observed vibrational response is due to both water molecules solvating protons and those in the bulk of the liquid, making an interpretation of the results in terms of structure highly challenging.

Recently, we have introduced the proton transfer mode as a sensitive probe to benchmark ultrafast dynamics of Zundel cations $H_5O_2^+$, prepared as the predominant species in the polar solvent acetonitrile.[33,34] Combining femtosecond 2D-IR and pump-probe[35] spectroscopy with quantum mechanics/molecular mechanics (QM/MM) simulations we have shown that the fluctuating electric field of the solvent induces a strong modulation of the proton double-minimum potential energy surface along the proton transfer coordinate $z$. Due to the absence of a central barrier in the double–well potential, the resulting stochastic large-amplitude displacements of the proton explore essentially all spatial positions along $z$ within 1 ps. Comparative experiments with protons in $H_2O$ reveal a strikingly similar behavior and suggest that the dimer (Zundel) motif occurs as a major solvation species in water. This conclusion has been confirmed in subsequent 2D-IR studies in a broader spectral range.[36]

In this Letter, we present a systematic study of proton dynamics in acetonitrile/water solvent mixtures by 2D-IR spectroscopy and QM/MM molecular dynamics simulations. In the experiments, the proton transfer mode is mapped at various acetonitrile/water mixing ratios from predominant acetonitrile to neat water, with heterogeneous acetonitrile/water solvation shells in the intermediate range. In parallel, the theoretical calculations are extended from the Zundel cation $H_5O_2^+$ in acetonitrile to geometries including up to 4 water molecules. We address the following key issues: How is the Zundel motif affected by the step-by-step addition of a solvating water environment? What is the influence of an increasingly aqueous environment on the dynamics of solvated protons? Which degrees of freedoms are relevant for proton motions at increasing water content?

Linear absorption spectra of one-molar (1 M) HI in mixtures of acetonitrile ($CH_3CN$) and water ($H_2O$) were recorded with an FTIR spectrometer in attenuated total reflection (ATR, Fig. 1d). HI fully dissociates in these solvents, resulting in a proton concentration of 1 M. In the measurements, the ratio of (relative) mole fractions $m_f(CH_3CN)/m_f(H_2O)$ was varied from 0.83 / 0.17 to 0.0 / 1.0 (neat $H_2O$). The residual relative fraction of 17% water in $CH_3CN$ originates from the water content of the aqueous HI stock solution. For each concentration ratio, a reference spectrum of the neat solvent mixture (not shown) was subtracted from the spectra in Fig. 1(d) to derive the spectra of solvated protons in Fig. 1(e).

The absorption spectrum of the Zundel cations in $CH_3CN$ displays the broad absorption band of the proton transfer mode between 900 and 1500 cm$^{-1}$, the HOH bending absorption of the flanking water molecules around 1730 cm$^{-1}$, the broad so-called Zundel continuum extending from approximately 1200 to 3000 cm$^{-1}$, and the OH stretching absorption of the flanking waters of the Zundel cation with maximum around 3380 cm$^{-1}$. With increasing water content, the proton transfer band undergoes a slight blue-shift (Figs. 1e, 2g) and a moderate decrease of the peak absorption by about 20%. The Zundel absorption continuum remains essentially unchanged upon changing the water content. The moderate spectral changes with increasing water fraction and the absence of new bands suggest the predominance of the Zundel motif in all cases, including neat $H_2O$.

Absorptive 2D-IR spectra for different mixing ratios of acetonitrile and water are presented in Fig. 2. All spectra were recorded with a waiting time T = 0 fs, results for longer waiting times of 50 fs and 100 fs are shown in the supporting information (SI, Fig. S1). Yellow-red contours in Fig. 2 represent a decrease of absorption due to ground state bleaching and stimulated emission on the fundamental (v = 0 to 1) transition of the proton transfer mode. The blue contours are due to a transient absorption increase on the v = 1 to 2 absorption. The zero crossing between the two components shows a slight blue-shift with increasing water content, similar to the linear proton transfer absorption (Fig. 2g). The 2D-IR spectra for all mixing ratios display the following common features: (i) the v = 1 to 2 absorption band is blue-shifted relative to the fundamental v = 0 to 1 transition, a hallmark of the double-minimum character of the proton potential along the proton transfer coordinate $z$.[33] (ii) All lineshapes are elongated along the excitation frequency axis $v_1$ due to the ultrafast spectral diffusion of the underlying vibrational excitations. This fact is illustrated in more detail by cuts of the 2D-IR spectra along the $v_1$ axis (Fig. S2 in the SI). (iii) The spectral positions of the 2D-IR bands are similar for the different $CH_3CN/H_2O$ mixing ratios, pointing to a similar double-minimum character of the proton potential along $z$, independent of the particular water content. These spectral signatures are distinctly different from the red-shifted v = 1 to 2 absorption of an anharmonic oscillator with a single minimum of the vibrational potential. The v = 1 state of the proton transfer mode decays with a sub-100 fs lifetime, followed by subpicosecond redistribution processes of the vibrational excess energy.[33,35,36]

The 2D-IR experiments were complemented by femtosecond pump-probe studies with resonant excitation of the proton transfer mode by pulses centered at 1210 cm$^{-1}$. Figure 1(f) shows transient pump-probe spectra of protons solvated in neat $H_2O$ (1 M HI), extending up to the onset of the OH stretching absorption of $H_5O_2^+$ around 3000 cm$^{-1}$. The absorption changes below 1600 cm$^{-1}$ are due to the reduced v=0 to 1 absorption and the enhanced v=1 to 2 absorption of the proton transfer mode, while the absorption changes at higher probe frequencies reflect anharmonic couplings to the other vibrations.[37] It is important to note that pump-probe signals are diminishingly small between 2400 and 2600 cm$^{-1}$, the range in which the OH stretching absorption band of the Eigen cation sets in.[35] We conclude that structural conversion of $H_5O_2^+$ into an Eigen form plays a minor role under the present experimental conditions. We further note that pump-probe spectra in the 2000-2600 cm$^{-1}$ range depend on the proton and counterion concentration as demonstrated in the SI for 2M HI (Fig. S3).

QM/MM molecular dynamics simulations were performed for protonated water clusters of increasing size embedded in a fluctuating acetonitrile environment, namely the Zundel cation $H_5O_2^+$, trimeric $H_7O_3^+$ and four water molecules with one excess proton ($H_9O_4^+$) (Figs. 1a-c). The water molecules and the excess proton were considered on the QM level while the acetonitrile environment was treated as MM region. The simulations follow the classical dynamics of the $H_xO_y^+$ nuclear degrees of freedom in long time dynamics and full coordinate space via an on-the-fly hybrid density functional theory treatment of electronic degrees of freedom (1.08 ns total simulation time, see SI for details). The on-the-fly treatment allows for, e.g., cluster dissociation, molecular rearrangements and transformation between Zundel- and Eigen-type species. The successive addition of water molecules mimics the experimental conditions where the acetonitrile-water ratio $m_f(CH_3CN)/m_f(H_2O)$ = 0.83 / 0.17 corresponds to ~3.5 water molecules per excess proton $H^+$ (Figs. 1, 2).

Results of the QM/MM simulations are summarized in Fig. 3. We find that the distribution of sampled O⋯O distances $R$ is gradually shifted towards larger values for an increasing number of $H_2O$ molecules (median value $R_m$ = 2.42, 2.50 and 2.57 Å for $H_5O_2^+$, $H_7O_3^+$ and $H_9O_4^+$). Despite the elongated *average* O⋯O distance, throughout the dynamics a designated $O_x$⋯$O_y$ pair exists that closely mirrors the O⋯O distance of $H_5O_2^+$ (see Fig. 3a for an excerpt of the $H_9O_4^+$ trajectory). By following the shortest O⋯O distance $R1$ (tagged via a 40 time-step moving average, cf. SI) we find a value R1 < 2.50 Å throughout predominant periods of the dynamics (~ 80 %) with only minor differences between $H_7O_3^+$ and $H_9O_4^+$ (median $R1_m$ = 2.46 Å and 2.47 Å, Fig. 3b). The second shortest O···O distance *R2* forms an overlapping but shifted distribution compared to *R1* and is almost indistinguishable for $H_7O_3^+$ and $H_9O_4^+$ (median $R2_m$ = 2.55 Å and 2.54 Å, Fig. 3c). The fourth water molecule in $H_9O_4^+$ is bound weaker (median $R3_m$ = 2.68 Å, Fig. 3d) and differences for Eigen-type configurations (*E*-$H_9O_4^+$, Fig. 1c) and wire configurations (*W*-$H_9O_4^+$, inset of Fig. 4b) can be identified (median $R3_m$ = 2.62 Å and 2.73 Å). Compared to *R1* and *R2*, the *R3* hydrogen bond strength in *W*-$H_9O_4^+$ is more comparable to the O…O hydrogen bond strength of liquid water.[38]

In Fig. 3e the distribution of explored proton transfer coordinate *z1* is compared for $H_5O_2^+$, $H_7O_3^+$ and $H_9O_4^+$. Due to the non-centrosymmetric surrounding of the active proton, asymmetric displacements are favored in $H_7O_3^+$ and $H_9O_4^+$ imposing a bimodal distribution in *z1*, while for $H_5O_2^+$ the distribution peaks around *z1* = 0. Nevertheless, the projection of sampled configurations on the two-dimensional space of the proton transfer coordinate *z1* and hydrogen bond coordinate *R1* (Fig. 3f) shows the strongly correlated dynamics of *R* and *z*[4,24,25] which are preserved for $H_7O_3^+$ and $H_9O_4^+$. The respective potential of mean force (PMF, SI Fig. S6) displays a prominent minimum centered around *R* = 2.42 Å and *z* = 0.0 Å ($H_5O_2^+$) while for $H_7O_3^+$ and $H_9O_4^+$ a small barrier, comparable to the thermal energy, arises at *z1* = 0.0 Å (188 cm$^{-1}$ and 242 cm$^{-1}$). Configurations indicative of the Eigen cation (*R* > 2.7 Å, see below for definition) are sampled infrequently during short ~ 50-100 fs periods of large excursions in the *R1* coordinate with slightly increased propensity for $H_7O_3^+$ and $H_9O_4^+$ compared to $H_5O_2^+$. Such structures are less stable than the potential minimum (907, 584 and 479 cm$^{-1}$ for $H_5O_2^+$, $H_7O_3^+$ and $H_9O_4^+$; Fig. S6). The standard deviation $\Delta R$ = ± 0.08 Å of the *R1* distribution is substantially affected by the acetonitrile solvent via a reshaping of the potential under the action of the fluctuating electric field. In summary, a picture for time averaged excess proton solvation structures in $H_7O_3^+$ and $H_9O_4^+$ emerges where a particularly strong O…O hydrogen bond *R1* is closely followed in strength by *R2*, together forming a central structural $H_7O_3^+$ unit. *R1* and *R2* are asymmetrically distorted due to the fluctuating solvent environment giving rise to a privileged dimer, characterized by $R1 < 2.50$ Å, that acts as IR chromophore and is existent throughout predominant periods of the dynamics.

The $R1/R2$ interconversion, i.e., the translocation of the $H_5O_2^+$ solvation species within strongly bound $H_7O_3^+$ is further analyzed in Fig. 4. For the correlation of the symmetrized proton transfer coordinates *(z1 + z2)/2* and the sum of hydrogen bond coordinates $R1 + R2$ we find that moderate collective compression of the $H_7O_3^+$ wire occurs (Fig. 4a for $H_9O_4^+$, data for $H_7O_3^+$, as well as $H_9O_4^+$ in *E*-$H_9O_4^+$ and *W*-$H_9O_4^+$ configurations are given in Fig. S7). The respective barrier for $R1/R2$ interconversion is ~ 350 cm$^{-1}$. The time scale of $R1/R2$ interconversion is further analyzed in Fig. 4b. We find that the histogram of $R1/R2$ interconversion times is modulated by O⋯O hydrogen bond motion at very early times and further shows an exponential decay with a characteristic time scale of ~ 150 fs. The motion of the strong O⋯O hydrogen bonds is characterized via the *<δR$_i$(t=0)δR$_i$(t)>* correlation functions of O⋯O elongations along *R* (Fig. 4c). The O⋯O period of $H_5O_2^+$ (dashed line) is reproduced for *R1* within *W*-$H_9O_4^+$ while the slower oscillations of *R2* and *R3* reflect the hierarchically weaker hydrogen bonds. Oscillations are damped on the 150-200 fs time scale and follow the decay of the electric field fluctuation correlation function imposed by the solvent environment.[33]

Our experimental and theoretical results demonstrate that proton dynamics for various acetonitrile-water ratios are governed by the fluctuations of the vibrational potential along the coordinates *z* and *R1-R3*, induced via electric field fluctuations from the surrounding solvent. The frequency spectrum of the fluctuating electric force is dominated by thermally activated solvent degrees of freedom, including fast librations. The latter are a source of the initial sub-100 fs decay in the correlation function of the fluctuating forces in both acetonitrile[39] and water.[22] For both solvents, the field fluctuation amplitudes projected on the proton coordinate *z* are in a range of 20-30 MV/cm and modulate the double minimum potential along *z* via the electronic polarizability.[19,20,33] The resulting stochastic motions along *z* are anharmonically coupled to the hydrogen bond coordinate and affect the O···O distances. In parallel, O···O excursions are triggered by the direct action of the fluctuating electric field. Both the proton transfer mode and the O···O vibrations respond with their intrinsic vibrational periods. The correlations between the nuclear motions decay on the time scale of the field fluctuation correlation function which is comparable to the time scale of the *R1/R2* interconversion in $H_7O_3^+$ and $H_9O_4^+$.

The 2D-IR spectra in Fig. 2 provide direct evidence for the predominance of a low-barrier double-minimum potential over the full range of acetonitrile-water mixing ratios. The 2D-IR data together with broadband pump-probe data (Fig. 1f) demonstrate the absence of a transformation between Zundel- and Eigen-type species within the observation time where any localization of the proton in Eigen-type species would result in a lasting change of the 2D-IR or pump probe spectra. Ultrafast fluctuations of the proton potential give rise to strong and ultrafast spectral diffusion of vibrational transitions, resulting in quasi-homogeneous 2D lineshapes between 1000 and 1300 cm$^{-1}$ which are established on a sub-100 fs time scale. In practically neat $CH_3CN$ the Zundel geometry is the by far dominating proton solvation structure.[33,34] The v = 1 to 2 absorption in Fig. 2a is blue-shifted relative to the v = 0 to 1 transition, a behavior which is characteristic for a double minimum potential with negligible barrier and fully reproduced by the QM/MM simulations (Fig. 3f). An increase of the water fraction in the solvent up to neat $H_2O$ changes the spectral positions and shape of the 2D-IR peaks to a minor extent only. With increasing water content, acetonitrile molecules in the solvation shell of the Zundel motif are replaced by water molecules which form strong-to-moderate hydrogen bonds with the flanking waters of $H_5O_2^+$. Such exchange shows negligible influence on the ultrafast fluctuation dynamics of the proton motions, i.e., the vibrational properties are determined by the $H_5O_2^+$ scaffold within a $H_7O_3^+$ unit.

The theoretical QM/MM results mimic the experimental conditions of ~3.5 water molecules per excess proton H$^+$ for the highest acetonitrile-water ratio. The simulations reveal a privileged dimeric $H_5O_2^+$ throughout the majority of the dynamics (80 %), characterized by $R1 < 2.50$ Å. This unit acts as the infrared chromophore via a harboring of the highly polarizable excess proton. This scenario is supported by the experimental observation (Figs. 1, 2) of a minor impact of the acetonitrile-water mixing ratio on the vibrational spectra, even for lowest water content. In $H_7O_3^+$ and $H_9O_4^+$ the privileged dimer is part of an asymmetrically distorted trimeric unit with a second O…O hydrogen bond that is slightly longer than in the dimer (2.54 - 2.55 Å vs. 2.46 - 2.47 Å). Compared to protonated water clusters in the gas phase[40] displacements in $R$ are increased (standard deviation $\Delta R = 0.08$ Å) which underscores a prominent role of solvent fluctuations on O…O dynamics. In the relevant range $R1 \pm \Delta R$, the central barrier of the proton double well potential is below the energy of the v = 0 state of the proton transfer coordinate $z$ (SI, Fig. S5).

The $R$-dependence of the $z$ ground state energy shows three distinct regions: (i) for $R < 2.48$ Å the relevant v = 0 quantum state is quasi unaffected by the vanishing barrier; (ii) for $2.48$ Å $< R < 2.7$ Å the barrier height becomes comparable to the energy of the v = 0 state. A sizeable central barrier in the proton potential arises for O…O distances $R > 2.56$ Å, that is expected to be further levelled by nuclear quantum effects[4] which suppress a persistent localization of the proton; (iii) localization occurs for $R > 2.7$ Å where the barrier exceeds the v = 0 energy, giving rise to Eigen-type species with a characteristic OH stretching absorption around 2665 cm$^{-1}$ [14,18].

The observed moderate blue shifts in linear absorption (Fig. 1) and transient 2D signals (Fig. 2) point to an increase in the fluctuation amplitude with increasing water content. For a neat water environment, the fluctuation amplitudes are expected to induce a moderate increase in the mean O…O distance $R$, in agreement with Ref. 4 (2.46-2.48 Å). This leads to a population increase of the transitional region ($R = 2.48$-$2.70$ Å) on the order of 10-20 % with a negligible increase of the population of Eigen–type species ($R > 2.7$ Å). Only extreme field fluctuation amplitudes (> 50 MV/cm) impose excursions $\Delta R \gg 0.1$ Å, that would lead to a barrier substantially exceeding the thermal energy. Due to the symmetry properties of the $H_5O_2^+$ complex, such extreme field values exist during very short periods only, and/or require a dedicated asymmetric surrounding with a preferential directionality of the external electric field. Our simulations on $H_7O_3^+$ and $H_9O_4^+$ show that even under such asymmetric conditions the O⋯O elongation of the $H_5O_2^+$ unit is moderate. We estimate that the high energy part of the potential occurring at $R > 2.7$ Å is populated for only ~9 % of the total time range covered by the calculated proton trajectories. The extremely rapid reshaping of the fluctuating potential suppresses a persistent localization of the proton in such a single-minimum well.

In the literature, the excursion $\delta = z$ of the proton position from its equilibrium average value $z = 0$ has been used as a parameter for distinguishing Zundel (small $\delta$) and Eigen geometries (large $\delta$).[4,25-27] Because of the ultrafast large-amplitude fluctuations of $\delta$ observed here, this quantity is not suitable for distinguishing structures within the ~1 ps lifetime of the Zundel motif. In other words, the time-dependent $\delta$ just mimics the large-amplitude excursions of the proton in the Zundel geometry rather than mapping a transition to a new solvation geometry of the proton (see also SI, Fig. S8). The proton excursions in the fluctuating Zundel motif are comparable to or even larger than proton translocations in a transition from a Zundel to Eigen geometry, i.e., there is no separation of the length scales of fluctuating motions and structure-changing steps. Thus, a physically and chemically meaningful parametrization of structures needs to distinguish the time range of stochastic proton motions from the comparably infrequent changes of solvation structure.

Our results are relevant for understanding the microscopic mechanism of proton translocation in aqueous solution via the von Grotthuss mechanism. The reported persistence of the Zundel motif strongly suggests a Zundel-type ground state during resting periods in-between proton translocation, in contrast to mechanistic proposals suggesting the Eigen-motif as a resting state.[25,26] The fluctuating proton's spatial amplitude within the Zundel species is comparable to the distance over which protons are shifted in an elementary transport step. This observation is in line with a picture in which a breaking of hydrogen bonds in the second solvation shell of $H_5O_2^+$ leads to a rearrangement of the hierarchical hydrogen bond structure, accompanied by translocation of the solvation geometry. Such rearrangements, imposed, e.g., via the angular jump mechanism[23], represent the key mechanism for inducing a transport step. Such events are clocked in picoseconds and define the lifetime of the Zundel species. It should be noted that the total $H_5O_2^+$ concentration remains unchanged within the ~1 ps time window of the present experiments, i.e., there is no net production of other proton solvation species.[33] Proton translocations between Zundel geometries leave the total absorption band due to the proton transfer mode unchanged and, thus, cannot be mapped here.

In conclusion, we have shown that the Zundel cation $H_5O_2^+$ represents a prominent proton solvation species in gradually changing polar solvent mixtures and neat water. During the lifetime of the $H_5O_2^+$ motif on the order of 1 ps, the proton undergoes fluctuating large-amplitude motions exploring essentially all possible positions between the flanking water molecules. Our in-depth theoretical simulations of proton solvation geometries including up to four water molecules in acetonitrile show that the dimeric Zundel motif is preserved and strongly interacts with its closest water neighbor in a $H_7O_3^+$ unit. The simulations further identify the range over which the proton transfer coordinate and the O⋯O hydrogen bond coordinate fluctuate. They suggest the absence of persistent proton localization within the lifetime of the Zundel motif.

**Acknowledgments.** B.P.F. gratefully acknowledges support through the Deutsche Forschungsgemeinschaft within the Emmy Noether Programme (grant FI 2034/1-1). This project has received funding from the European Research Council (ERC) under the European Union's Horizon 2020 research and innovation programme (grant agreement No 802817).

**Supporting information available**: Experimental methods and data analysis, 2D infrared spectra, theoretical methods.

**Figure Captions**

**Figure 1**. Schematic structures of (a) the Zundel cation $H_5O_2^+$, (b) the trimeric water cluster $H_7O_3^+$, and (c) a cluster of four water molecules with one excess proton ($H_9O_4^+$) in Eigen-type, star-like configuration ($E$-$H_9O_4^+$). The symbols z represent the proton/hydrogen stretching coordinates, the symbols R the O...O distances. (d) Linear infrared absorption spectra of 1 M of 67% HI in acetonitrile/water ($CH_3CN/H_2O$) mixtures with the relative mole fractions $m_f$ indicated. The spectra were recorded with an FTIR spectrometer in Attenuated Total Reflection (ATR) mode and display bands of solvated protons and the solvents. (e) Difference ATR spectra $\Delta OD$ of solvated protons after subtraction of the solvent absorption for the different mixtures. The negative $\Delta OD$ in the frequency range around 3300 cm$^{-1}$ are caused by the subtraction of the OH stretch absorption of water molecules not being part of the immediate proton solvation structure. The narrow features in the difference spectra are due to the imperfect subtraction of very strong and narrow acetonitrile bands. (f) Transient femtosecond pump-probe spectra (colored lines) of protons in neat water (1 M of 67% HI) recorded after excitation with pump pulses centered at 1210 cm$^{-1}$ (dash-dotted line: pump spectrum). The thinner black line represents the linear absorption spectrum (cf. Fig. 1e).

**Figure 2**. (a-f) 2D-IR spectra of the proton transfer mode in a series of acetonitrile/water mixtures with a relative acetonitrile fraction decreasing from (a) 0.83 to (f) 0.0. The absorptive 2D signal recorded at a waiting time T=0 fs is plotted as a function of the excitation frequency $\nu_1$ and the detection frequency $\nu_3$. Yellow-red contours represent an absorption decrease, blue contours an absorption increase. Amplitudes are scaled relative to the maximum absorption decrease, the signal change between neighboring contour lines is 10 %. (g) Frequency position of the absorption maximum in the linear proton transfer band (red symbols, cf. Fig. 1e). The black symbols represent the frequency position of the zero crossing between the absorption decrease and increase in the 2D-IR spectra.

**Figure 3**. Results of QM/MM molecular dynamics simulations: (a) time evolution of O⋯O distances $R$ in the $H_9O_4^+$ trajectory (dashed lines) together with the 40 time-step moving average (solid lines). (b-d) Distribution of sampled O⋯O distances $R$, as well as shortest to third shortest O⋯O distance $R1$-$R3$ for investigated clusters $H_5O_2^+$, $H_7O_3^+$ and $H_9O_4^+$. (e) Distribution of proton transfer coordinate $z1$ in $H_5O_2^+$, $H_7O_3^+$ and $H_9O_4^+$. (f) Sampled configurations in two-dimensional space of the proton transfer coordinate z1 and hydrogen bond coordinate $R1$ in $H_5O_2^+$ and $H_9O_4^+$.

**Figure 4**. (a) Correlation of the symmetrized proton transfer coordinates *(z1 + z2)/2* and the sum of hydrogen bond coordinates *R1 + R2* of $H_9O_4^+$ (data for $H_7O_3^+$, *E*-$H_9O_4^+$ and *W*-$H_9O_4^+$ configurations given in Fig. S7). (b) Time scale of *R1*/*R2* interconversion in $H_9O_4^+$ with exponential fit (~150 fs decay) indicated as solid line. The insert illustrates *R1*/*R2* interconversion in *W*-$H_9O_4^+$. (c) Correlation functions of O⋯O elongations along $R$ *<δR$_i$(t=0) δR$_i$(t)>* in *W*-$H_9O_4^+$ with i = 1,2,3. The black dashed line indicates the time scale of cluster disintegration.

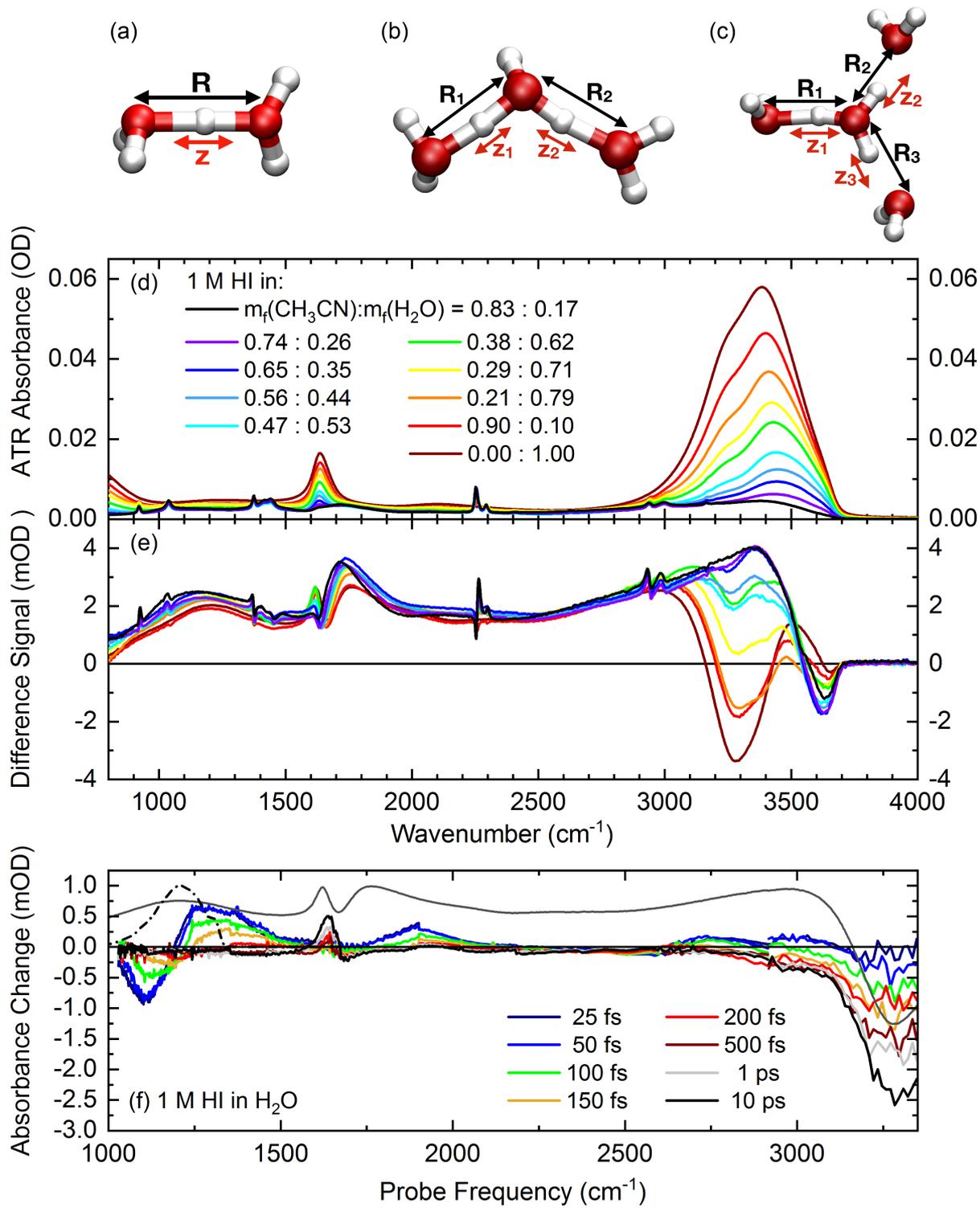

Figure 1

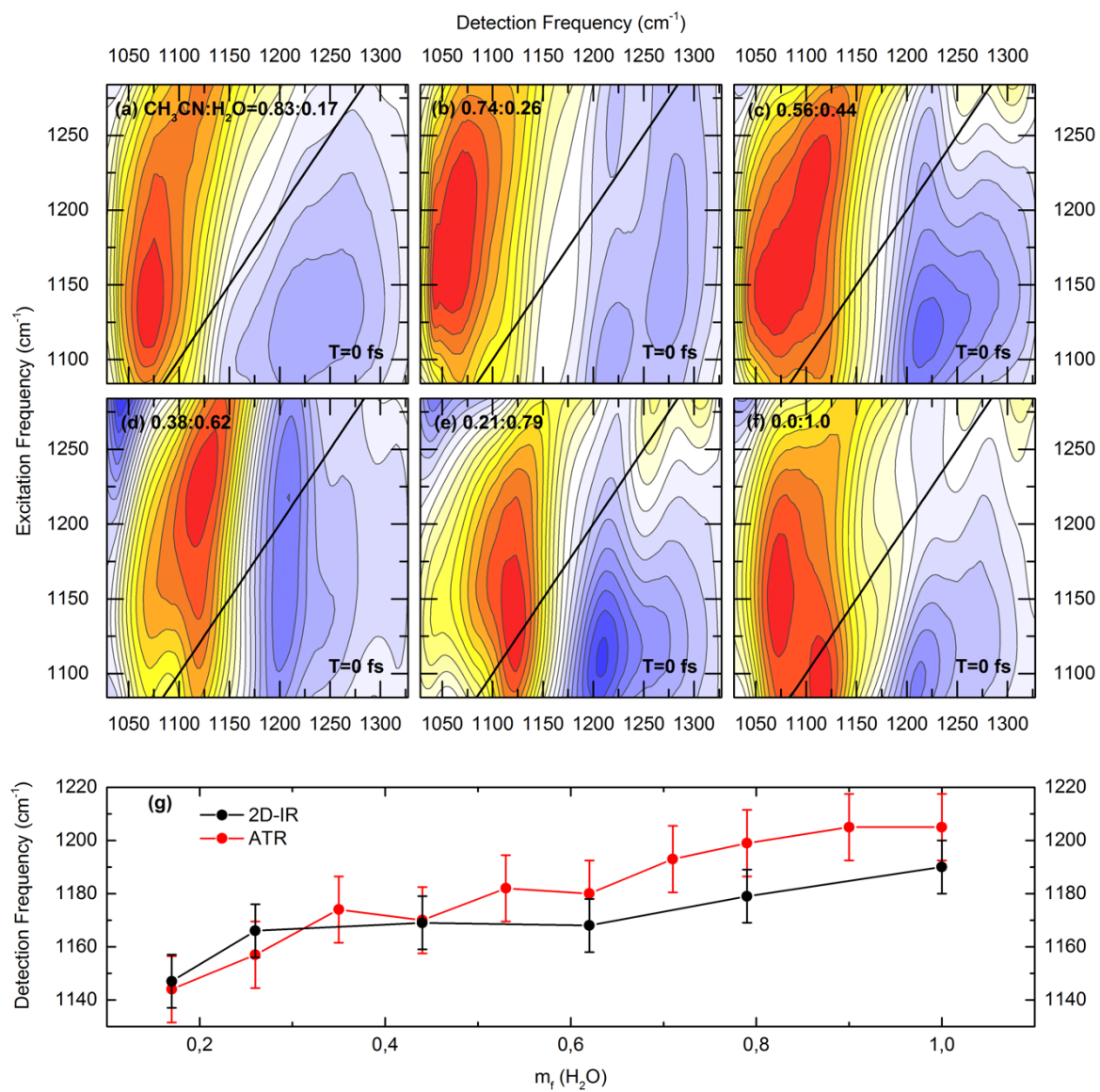

Figure 2

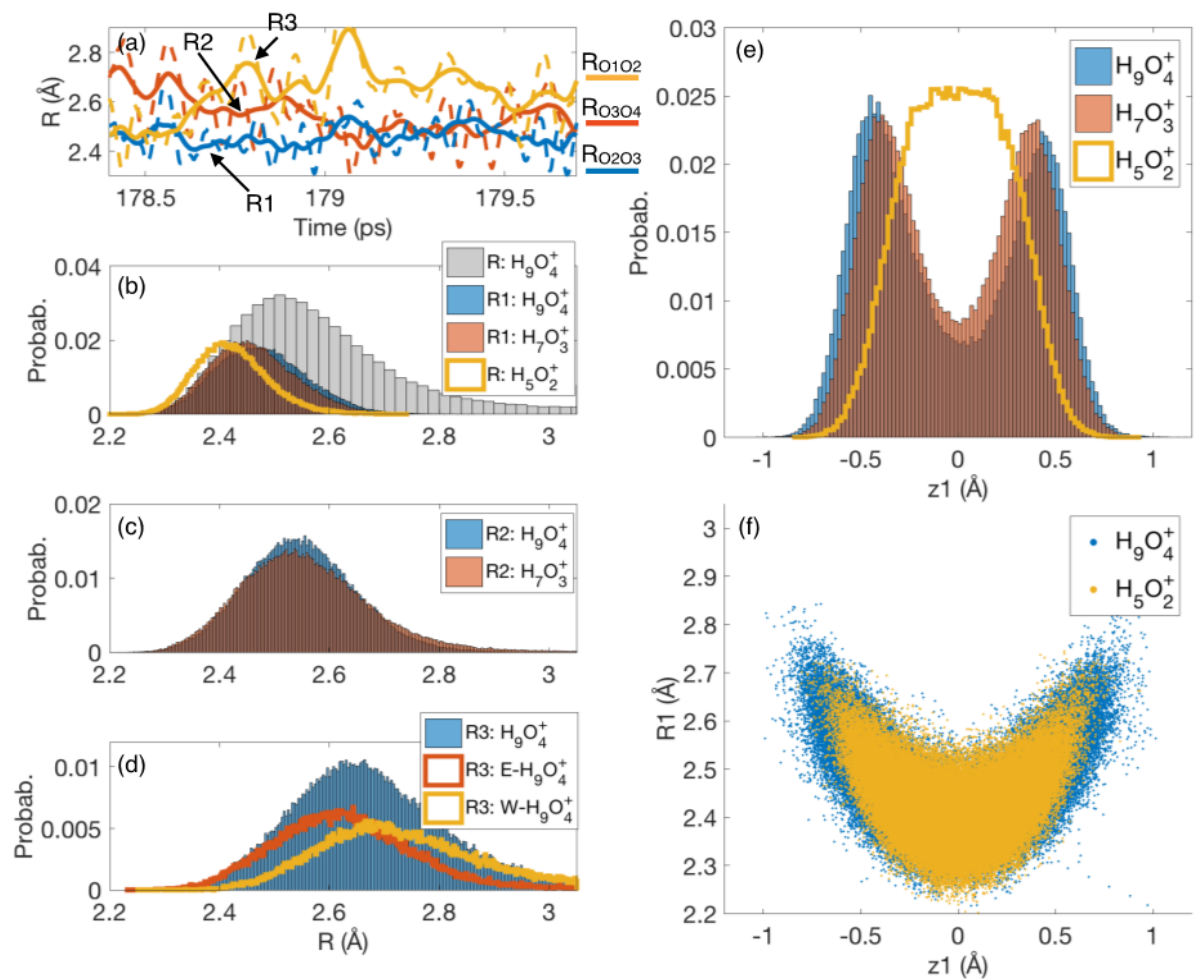

Figure 3

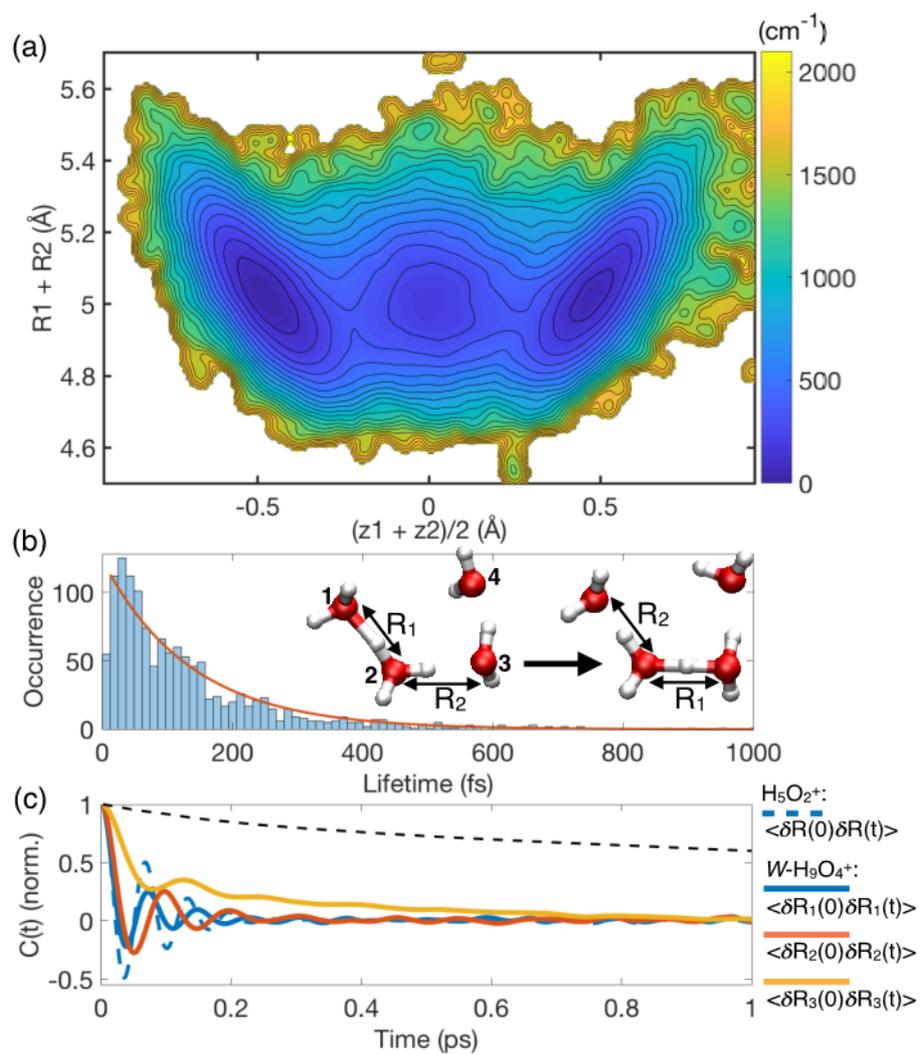

Figure 4